\documentclass[twocolumn,showpacs,preprintnumbers,amsmath,amssymb,epsfig,widetext]{revtex4}

\usepackage{graphicx}
\usepackage{dcolumn}
\usepackage{bm}
\usepackage{epsfig}

\input epsf

\def\be{\begin{equation}}
\def\ee{\end{equation}}
\def\ba{\begin{eqnarray}}
\def\ea{\end{eqnarray}}

\newcommand{\potential}{{A}}
\newcommand{\shift}{{B}}
\newcommand{\curvature}{{H}_L}
\newcommand{\shear}{{H}_T}
\newcommand{\kh}{k_H}

\begin{document}

\preprint{}

\title{A Parameterized Post-Friedmann Framework for Modified Gravity}

\author{Wayne Hu$^{1,2}$ and Ignacy Sawicki$^{1,3}$}
\email{whu@background.uchicago.edu}
\affiliation{{}$^1$ Kavli Institute for Cosmological Physics, Enrico Fermi
Institute,  University of Chicago, Chicago IL 60637 \\
{}$^2$ Department of Astronomy \& Astrophysics,  University of Chicago, Chicago IL 60637\\
{}$^3$ Department of Physics,  University of Chicago, Chicago IL 60637
}

\date{\today}

\begin{abstract}
We develop a parameterized post-Friedmann (PPF) framework which
describes three regimes of modified gravity models that
accelerate the expansion without dark energy.   On large scales, the evolution of 
scalar metric
and density perturbations must be compatible with the expansion history defined by
distance measures.  On intermediate scales in the linear regime, 
they form a scalar-tensor theory with a modified Poisson equation.  On small scales in 
dark matter halos such as our own galaxy, modifications  must
be suppressed in order to satisfy stringent local tests of general relativity.  We 
describe these regimes with three free  functions
and two parameters: the relationship between the two metric fluctuations, 
the large and intermediate scale relationships to density fluctuations and the two scales
of the transitions between the regimes.  We also clarify the formal equivalence of
modified gravity and  generalized dark energy. 
The PPF description of linear fluctuation in  $f(R)$ modified action and the Dvali-Gabadadze-Porrati
braneworld models show excellent agreement with explicit calculations.  
Lacking cosmological simulations of
these models, our non-linear halo-model description remains an ansatz but
one that enables well-motivated consistency tests of general relativity.  
The required suppression of modifications within  dark matter
halos suggests that the linear and weakly non-linear regimes are better suited for making
complementary test of general relativity than the deeply non-linear regime.  
\end{abstract}


\maketitle

\section{Introduction}
\label{sec:introduction}

Theoretically compelling alternatives to a cosmological constant as the source of the
observed cosmic acceleration are currently lacking. 
In the absence of such alternatives, it is useful to
have a phenomenological parameterized approach for testing the predictions of a
cosmological constant and phrasing constraints in a model-independent
language.   This approach parallels that of local tests of general relativity.
The parameterized post-Newtonian description of gravity forms a complete description
of leading order deviations from general relativity locally under a well-defined set
of assumptions \cite{Will:2005va}.  

A parameterization of cosmic acceleration from the standpoint of dark energy is
now well-established.   The expansion history that controls distance observables
is completely determined by the current dark energy density and its equation of
state as a function of redshift.   Structure formation tests involve additional parameters
that control inhomogeneities in the dark energy.  Covariant conservation of 
energy-momentum
requires that the dark energy respond to metric or gravitational potential fluctuations
at least on scales above the horizon.   
In a wide class of models where the dark energy
remains smooth relative to the matter on small scales, the phenomenological parameter
of interest is where this transition occurs \cite{CalDavSte98,Hu98,Hu04c}.

A similar structure is imposed on modified gravity models that accelerate
the expansion without dark energy.  
Requirements that gravity remain a metric theory where energy-momentum
is covariantly conserved also place strong constraints their scalar
degrees of freedom.   On
scales above the horizon, structure evolution must be compatible with the
background expansion \cite{Ber06}.   Intermediate scales are characterized by 
a scalar-tensor theory with a modified Poisson equation \cite{LueScoSta04}.   
If these modifications
are to pass stringent local tests of gravity then additional scalar degrees of
freedom must be suppressed locally \cite{DefDvaGabVai02}.   
Two explicit models that exhibit all three 
regimes of modified gravity are the so-called $f(R)$ modified Einstein-Hilbert 
action models \cite{Caretal03,NojOdi03,Capozziello:2003tk}
and the Dvali-Gabadadze-Porrati (DGP) braneworld model \cite{DvaGabPor00}.  

Although several parameterized gravity approaches exist in the literature,
none describe all three regimes of modified gravity (cf.~\cite{CalCooMel07,Amendola:2007rr,ZhaLigBeaDod07})
and most do not explicitly enforce a metric structure to gravity or
energy momentum conservation 
 (e.g.~\cite{IshUpaSpe06,KnoSonTys06,WanHuiMayHai07,Son06,HutLin07}).

In this paper, we develop a parameterized post-Friedmann (PPF) framework that
describes all three regimes of  modified gravity models  that
accelerate the expansion without dark energy.  We begin in \S \ref{sec:trinity} by
describing the three regimes individually and the requirements they impose on the
structure of such modifications.  In \S \ref{sec:linear}, we describe a linear theory
parameterization of the first two regimes and test it against explicit calculations
of the $f(R)$ and DGP models.  In \S \ref{sec:nonlinear}, we develop a non-linear
ansatz for the third regime based on the halo model of non-linear clustering.
In the Appendix, we clarify the formal relationship between modified gravity and
dark energy beyond the smooth class of models.

\section{Three Regimes of Modified Gravity}
\label{sec:trinity}

In this section, we discuss the three regimes of modified gravity theories
that accelerate the
expansion without dark energy.   We begin by 
 reviewing the requirements on super-horizon metric perturbations
  imposed by compatibility with a given expansion history (\S \ref{sec:superhorizonregime}).
  Modifications that introduce extra scalar degrees of freedom then enter
  a quasi-static regime characterized 
  by a modified Poisson equation (\S \ref{sec:quasistaticregime}).  
  Finally stringent local tests of gravity require that modifications
  are suppressed in the non-linear regime within collapsed dark matter haloes
  (\S \ref{sec:nonlinearregime}).

\subsection{Post-Friedmann Super-horizon Regime}
\label{sec:superhorizonregime}

Under the assumption that modified gravity remains a metric theory in a statistically homogeneous and isotropic cosmology
 where energy-momentum is covariantly conserved, 
 a parameterization of the expansion history that is complete under general
relativity is complete under modified gravity as well.   A modified gravity model
and a dark energy model  with
the same expansion rate $H= a^{-1} da/dt$ and
spatial curvature predicts the same observables for any measure that
is based on the distance-redshift relation (see e.g.~\cite{Teg01}).
 Hence in terms of the background, modified gravity
models can be parameterized in the same way as dark energy without
loss of generality.  
Neglecting spatial curvature and radiation for simplicity here and throughout, 
we can assign an effective energy density 
\begin{equation}
\rho_{\rm eff} = {3 \over 8\pi G} ( H^2  - H_m^2 a^{-3}) \,,
\end{equation}
where 
\begin{equation}
H_m^2 \equiv {8\pi G \over 3} \rho_m(\ln a=0)
\end{equation}
would be the contribution of matter to the expansion under the normal
Friedmann equation.    Alternately we can assign a current effective
energy density in units of the critical density 
$\Omega_{\rm eff} =  8\pi G\rho_{\rm eff}/3 H_0^2$ and an effective
equation of state
\begin{equation}
1+w_{\rm eff}(\ln a) \equiv - {1\over 3}{\rho_{\rm eff}' \over \rho_{\rm eff}}
=-{1 \over 3}{2H H' + 3 H_m^2 a^{-3} \over H^2 - H_m^2 a^{-3}} \,.
\end{equation}  

Compatibility with  this expansion
combined with energy momentum conservation
highly constrains the evolution of metric fluctuations
above the horizon.  Super-horizon metric fluctuations
in a perturbed universe can be viewed as evolving as a separate
 universe
 under the same modified Friedmann equation but with different parameters.

Bertschinger \cite{Ber06} showed that consequently  metric
fluctuations in fact obey the same fundamental constraints as they do in general relativity.  
These constraints appear in different ways in different gauges as detailed in the Appendix.
Under the assumptions of a metric theory and energy-momentum conservation, 
all of the usual gauge structure including the so-called ``gauge invariant" approach
used here
apply to modified gravity as well.  

In the comoving gauge of the matter, the constraint for adiabatic initial conditions
looks particularly simple.   The curvature or 
space-space piece of the metric fluctuation $\zeta$ remains constant to leading order  
(\cite{Bar80}, see also (\ref{eqn:zetaprimeeff}))
\begin{equation}
\zeta ' = {\cal O}(\kh^{2}\zeta) \,,
\label{eqn:zetaconservation}
\end{equation}
where $' = d/d\ln a$ and $\kh = k/aH$ is the wavenumber in units
of the Hubble parameter.
In the more familiar Newtonian gauge where the curvature is
denoted $\Phi$ and the time-time piece or gravitational potential
$\Psi$, the gauge transformation equation~(\ref{eqn:comnewt})
\begin{equation}
\zeta = \Phi - V_{m}/\kh 
\label{eqn:zetaphi}
\end{equation}
and the momentum conservation equation~(\ref{eqn:conservation})
\begin{equation}
V_m' + V_m = \kh \Psi  
\end{equation} 
along with Eqn.~(\ref{eqn:zetaconservation}) imply
\begin{equation}
\Phi'' - \Psi'  -{H'' \over H'}\Phi' - \left( {H' \over H} -{H'' \over H'} \right) \Psi =
 {\cal O}(\kh^{2}\zeta) \,.
 \label{eqn:bertschinger}
\end{equation}
Here $V_{m}$ is the scalar velocity fluctuation  of the matter in both
the comoving and Newtonian gauges (see Eqn.~(\ref{eqn:stressenergy}) and
Eqn.~(\ref{eqn:fluidtrans})).    Eqn.~(\ref{eqn:bertschinger})
is also satisfied in general relativity \cite{HuEis99}.

These relations have been explicitly shown to hold for
DGP braneworld gravity \cite{SawSonHu06} and $f(R)$ modified action gravity
\cite{SonHuSaw06}.
What distinguishes a particular model of gravity or dark energy is the relationship 
between $\Phi$ and $\Psi$ in the Newtonian gauge or equivalently
$\zeta$ and $V_{m}$ in the comoving gauge.    
We will parameterize this relation by
\begin{equation}
g \equiv {\Phi +\Psi \over \Phi -\Psi} = {\kh \zeta + V_m' + 2 V_m \over \kh \zeta - V_m'}\,.
\label{eqn:gdefinition}
\end{equation}
In terms of the post-Newtonian parameter $\gamma = -\Phi/\Psi$, 
$g= (\gamma-1)/(\gamma+1)$.  Given that gravitational redshift and lensing
effects involve the metric combination
\begin{equation}
\Phi_{-} = {\Phi -\Psi \over 2} \,,
\end{equation}
we will typically state metric results in terms of $\Phi_-$.  It is useful to note
that $\Phi = (g+1)\Phi_-$ and $\Psi = (g-1)\Phi_-$.
Super-horizon scalar metric fluctuations for adiabatic perturbations are
completely defined by the expansion history $H$ and the metric
ratio $g$. 

\begin{figure}[tb]
\begin{center}
\includegraphics[width=3.4in]{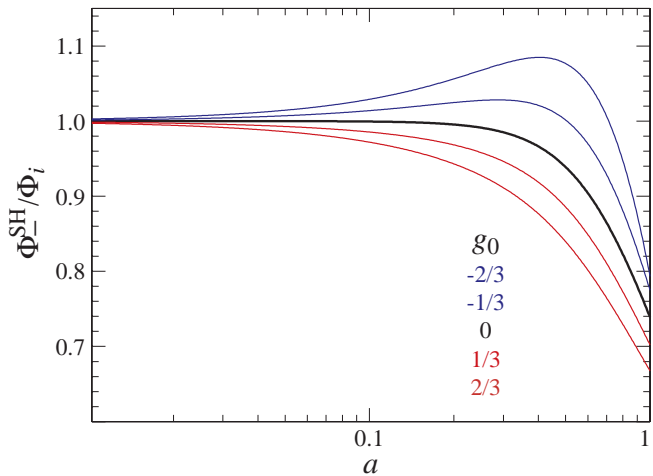}
\caption{Super-horizon (SH) metric evolution of $\Phi_{-}=(\Phi-\Psi)/2$
for various choices of $g=g_{0}a$.
For $g<0$, $\Phi_{-}$ can actually grow near the onset of acceleration.
The expansion history is fixed by
$w_{\rm eff}=-1$ and $\Omega_{m}=0.24$ in all cases. } \label{fig:timesh}
\end{center}
\end{figure}

 In Fig.~\ref{fig:timesh} we show the evolution of $\Phi_-$
given a metric ratio that evolves as $g = g_0 a$ for a $\Lambda$CDM expansion history 
defined by $w_{\rm eff}=-1$ and $1-\Omega_{\rm eff}=\Omega_m= 0.24$.  Given that $g \rightarrow 0$
as $a \rightarrow 0$, we take matter dominated
initial conditions of $\Phi_- = \Phi_i = 3\zeta_i/5$ and $\Phi_-'=0$ as in general
relativity.  Note that if $g < 0$, $\Phi_-$ can actually grow during the acceleration
epoch.  
Changing  the evolution of gravitational potentials alters the
low order multipoles of the CMB through the integrated Sachs Wolfe effect.  For
example it can be used to suppress the CMB quadrupole (see e.g. \cite{SonHuSaw06}).

Under general relativity,
the metric ratio is determined by the ratio of anisotropic stress to energy density 
(see Eqn.~(\ref{eqn:phiplusgr})).
 It supplements the information in the expansion history since in the
background the anisotropic stress vanishes by assumption.  
In particular, with the matter carrying negligible anisotropic stress, this relationship
is determined by the anisotropic stress of the dark energy.  For models of dark
energy based on scalar fields, this also vanishes in linear theory
and hence $g=0$.  Nevertheless, a
hypothetical dark energy component that produces
the same $g$ as a modified gravity model cannot be distinguished from super-horizon metric
fluctuations.  We further examine the relationship between modified gravity and
dark energy in the Appendix.


\subsection{Post-Newtonian Quasi-Static Regime}
\label{sec:quasistaticregime}

Well inside the horizon but still in the regime of linear fluctuations, 
theories that exhibit an extra scalar degree of freedom tend toward a
post-Newtonian scalar-tensor description.  As is the case in 
general relativity when time derivatives of the metric fluctuations can be
ignored compared with spatial gradients, the modified field equations 
reduce to a modified Poisson equation 
\begin{equation}
k^2 \Phi_- = {4\pi G \over 1+f_G} a^2 \rho_m \Delta_{m} \,,
\label{eqn:modpoisson}
\end{equation}
where $\Delta_m$ is the fractional density perturbation (in the matter comoving gauge), 
$f_G$ parameterizes a possibly time-dependent modification of the Newton constant,
and the relationship between the two metric fluctuations is again parameterized
by $g$ as in Eqn.~(\ref{eqn:gdefinition}). 
We assume that spatial fluctuations in $f_G$ lead to a second order correction
in the linear regime.
 We call this the quasi-static approximation.
This quasi-static approximation has been explicitly shown to hold for
both  DGP braneworld gravity \cite{KoyMaa06,SawSonHu06} 
and $f(R)$ models \cite{Zha06,SonHuSaw06}
once $\kh \gg 1$.  

\begin{figure}[tb]
\begin{center}
\includegraphics[width=3.4in]{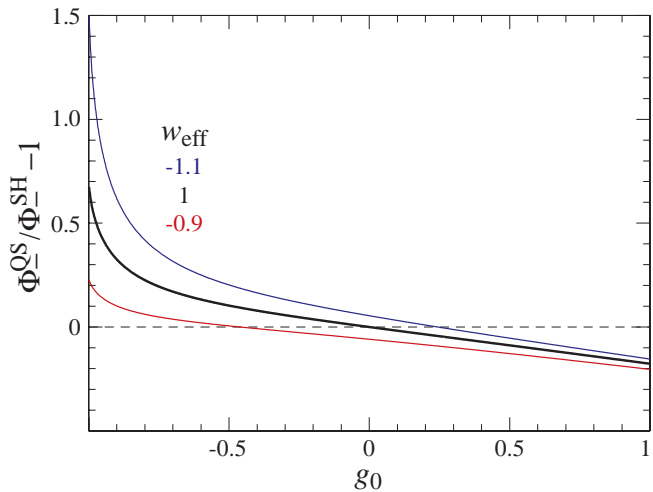}
\caption{Fractional difference between the quasi-static (QS) vs super-horizon (SH) metric at $a=1$ 
for a metric ratio of $g=g_0 a$.
The two evolutions differ unless $w_{\rm eff}=-1$ and $g_{0}=0$.  $\Omega_{m}=0.24$ for
all cases.} \label{fig:shqs}
\end{center}
\end{figure}

Density fluctuations on the other hand are determined by 
the Newtonian limit of the conservation equations
\begin{align}
\Delta_m' &= -\kh V_m \,, \nonumber\\
V_m' + V_m &= \kh \Psi  = (g-1)\kh \Phi_{-}\,.
\label{eqn:newtonianconservation}
\end{align}
Combining the modified Poisson equation (\ref{eqn:modpoisson}) and the 
conservation equations (\ref{eqn:newtonianconservation}) and taking $f_G=$const.
for simplicity yields
\begin{align} 
& \Phi_-'' + \left( {4 + {H' \over H} } \right) \Phi_-'\\ & \qquad + \left[
{3 + {H' \over H} + {3 \over 2}{H_m^2 \over (1+f_G) H^2 a^3}
(g-1) } \right] \Phi_- =0  \,.\nonumber
\end{align}
Note that this quasi-static equation is inequivalent to the super-horizon evolution
(\ref{eqn:bertschinger}) whenever $w_{\rm eff} \ne -1$, $g \ne 0$,  or $f_G\ne 0$.  
The differences for $w_{\rm eff}\ne -1$ applies in general relativity as well and corresponds
to the well-known fact that growth defined by a smooth dark energy component is
inconsistent with conservation of energy momentum on super-horizon scales 
(e.g. \cite{CalDavSte98,Hu98}).   In Fig.~\ref{fig:shqs}, we compare the quasi-static
and super-horizon evolution to the present.    Note that in terms of  the change in
the gravitational potential from its initial value, important for gravitational redshift
effects in the CMB, the differences are amplified.  Compared with $\Lambda$CDM where
$\Delta \Phi_- \approx -\Phi_i/4$ (see Fig.~\ref{fig:timesh}), these changes are enhanced by a factor of $\sim 4$ and
cannot be neglected for large values of $|g|$.

\subsection{General Relativistic Non-linear Regime}
\label{sec:nonlinearregime}

A successful modification of gravity must have a third regime where non-linearities
in the modified field equations bring the dynamic back to general relativity. 
Scalar-tensor modifications of gravity that persist to small scales can be ruled
out by stringent local tests of gravity.   For example, the Cassini mission imposes the limit
\cite{Will:2005va}
\begin{equation}
|g| < 1.2 \times 10^{-5} 
\end{equation}
for the metric ratio in the solar system.

In both the DGP braneworld and $f(R)$ modifications, the extra scalar degree of
freedom obeys a non-linear equation that suppresses its effects within collapsed
objects such as dark matter halos.  
In the DGP braneworld model, the scalar degree of freedom represents a brane
bending mode.  Non-linear interactions of this mode become important at the
so-called Vainshtein radius $r_* \sim (2G M r_c^2 )^{1/3}$ \cite{DefDvaGabVai02}.  
Here $r_c \sim H_0^{-1}$
is the crossover scale where gravity becomes fully 5D.  This radius is comparable
to the virial radius of dark matter halos and suggests that inside a halo
gravitational interactions behave as in general relativity.  
 
Similarly, for modified action $f(R)$ models the extra scalar degree of
freedom corresponds to $d f/dR$ and 
has a mass that depends on the local curvature $R$.   
Deviations from general relativity can then be suppressed
by the so-called chameleon mechanism \cite{KhoWel04,Mota:2003tm}
 so long as the gravitational potential
is sufficiently deep \cite{FauTegBun06,HuSaw07}.  

In the empirical PPF description that follows
in \S \ref{sec:linear} and \S \ref{sec:nonlinear} we seek a description that 
joins these three regimes.

\section{Linear Theory Parameterization}
\label{sec:linear}
 
 In this section, we construct a parameterized framework for linear perturbations
 in modified gravity models that joins the super-horizon and quasi-static regimes
 described in the previous section.   We first describe the construction  
(\S \ref{sec:ppfparameters}) and then test the parameterization against
 modified action $f(R)$ models (\S \ref{sec:frlinear}) and the DGP braneworld
 model (\S \ref{sec:dgplinear}).

\subsection{PPF Parameters}
\label{sec:ppfparameters}

We have seen in \S \ref{sec:trinity} that in both the super-horizon and quasi-static
regimes, the evolution of metric fluctuations are primarily determined by the metric ratio
$g$.   However, the manner in which the metric ratio $g$ determines metric evolution
differs between the two regimes.    The quasi-static regime also allows the freedom to change the effective Newton
constant.  

Let us introduce a parameterization that bridges the dynamics of the two regimes
at a scale that is parameterized in units of the Hubble scale. 
The Newtonian-limit conservation equations (\ref{eqn:newtonianconservation})
must first be corrected for metric evolution.  The
exact conservation equations imposed by $\nabla^{\mu} T_{\mu \nu}=0$ and the
metric is given by (see Eqn.~(\ref{eqn:energycom}) and (\ref{eqn:momentumnewt}))
\begin{align}
\Delta_m' &= -\kh V_m - 3\zeta' \,, \nonumber\\
V_m' + V_m &= ( g-1)\kh   \Phi_- \,, 
\label{eqn:comconservation}
\end{align}
where the additional term involving  evolution of the metric is given in the matter comoving gauge $\zeta'$
to match the definition of density perturbations $\Delta_m$ in this gauge.   It is related to the
evolution of the Newtonian metric by Eqn.~(\ref{eqn:zetaphi})
\begin{align}
\zeta' &=  (g+1)\Phi_-'  + (1-g+g')\Phi_- -  {H' \over H} { V_m \over \kh} \,.
\label{eqn:zetaprimedefn}
\end{align}
In order to match the
super-horizon scale behavior we introduce an additional term 
 $\Gamma$ to the modified Poisson equation~(\ref{eqn:modpoisson})
 \begin{equation}
k^2 [\Phi_- + \Gamma] = 4\pi G a^2 \rho_m\Delta_m \,.
\label{eqn:gammapoisson}
\end{equation}
We now demand that as $\kh \rightarrow 0$ $\Gamma$ enforces the metric evolution of
Eqn.~(\ref{eqn:bertschinger}).  In this limit,
the derivative of Eqn.~(\ref{eqn:gammapoisson})
 gives an evolution equation for $\Gamma$ given the
conservation equations (\ref{eqn:comconservation}) and the required metric evolution.  Aside
from $g$, the only remaining freedom is determining the leading order behavior of $\zeta'$.  Without loss of generality, we can parameterize Eqn.~(\ref{eqn:zetaconservation})
with a possibly time-dependent function $f_\zeta$
\begin{equation}
\lim_{\kh \rightarrow 0} \zeta' 
 = {1 \over 3}f_\zeta \kh V_m \,.
\end{equation}
Although the super-horizon metric is determined by $H$ and $g$ alone, its relationship
to the comoving density perturbation is not.  Since $\kh V_m = {\cal O}( \kh^2\zeta)$ and
$\zeta'=   {\cal O}( \kh^2\zeta)$, this degree of freedom enters into the conservation 
equation (\ref{eqn:comconservation}) at leading order.
Combining these relations, we obtain the equation of motion for $\Gamma$ 
 \begin{align}
 \Gamma' +\Gamma  
&=S \,,  \qquad (\kh \rightarrow 0)\,, 
\end{align}
where the source is
\begin{align}
S &= - {\left[  {1\over g+1}{H' \over H}  + {3 \over 2}{H_{m}^{2}\over H^{2}a^{3}} (1+f_\zeta)\right] 
 } {V_m \over \kh}
\nonumber\\&  \qquad + \left[ { g' -2 g \over g+1} \right]
 \Phi_{-} \,.
\end{align}
Here we have kept only the leading order term in $\kh$.

Note that the exact choice of $f_\zeta$ is rarely important for observable quantities.   
Any choice will produce
the correct behavior of the metric evolution since that depends only on enforcing
$\zeta'=   {\cal O}( \kh^2\zeta)$.  Hence observables associated with gravitational redshifts
and lensing are not sensitive to this choice.    Only observables that depend on the
comoving density on large scales beyond the quasi-static regime
are affected by this parameter.   Furthermore the super-horizon density perturbation in Newtonian gauge
or any gauge where the density fluctuation evolves as the metric fluctuation is also
insensitive to $f_\zeta$.

On small scales, recovery of the modified Poisson equation~(\ref{eqn:modpoisson}) from 
(\ref{eqn:gammapoisson}) implies
\begin{equation}
\Gamma = f_{G}\Phi_{-}\,,  \qquad (\kh \rightarrow \infty) \,.
\end{equation}
Finally to interpolate between these two limits we take the full equation of motion for $\Gamma$ to be 
 \begin{equation}
( 1+ c_\Gamma^2 \kh^{2})
\left[ \Gamma' +\Gamma + 
c_\Gamma^2 \kh^{2 }\left(\Gamma-f_G\Phi_-\right)
\right]= S \,.
\label{eqn:gammaeom}
\end{equation}
For models where $S \rightarrow 0$ as $a \rightarrow 0$ we take initial conditions of
$\Gamma=\Gamma'=0$ when the mode was above the horizon.

In summary, given an expansion history $H(a)$,
our PPF parameterization is defined by 3 functions and 1 parameter: 
the metric ratio $g(\ln a,\kh)$,
the super-horizon relationship between the metric and density $f_\zeta(\ln a)$,  the quasi-static
relationship or scaling of Newton constant $f_G(\ln a)$, and the relationship between the
transition scale and the Hubble scale $c_\Gamma$.    For models which modify gravity
only  well after matter radiation equality, these relations for
the metric, density and velocity evolution combined with the usual transfer functions completely
specify the linear observables of the model.  In specific models, these functions can 
themselves be
simply parameterized as we shall now show for the  $f(R)$ and DGP models.

\subsection{$f(R)$ Models}
\label{sec:frlinear}

\begin{figure}[tb]
\begin{center}
\includegraphics[width=3.4in]{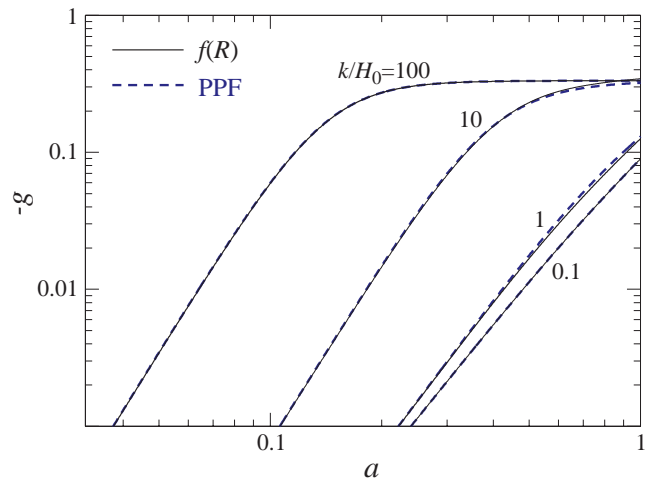}
\caption{Evolution and scale dependence of the metric ratio
$g$ in $f(R)$ models compared
with the PPF fit.  Here $B_{0}=0.4$, $w_{\rm eff}=-1$ and $\Omega_{m}=0.24$.}
 \label{fig:gfit_fr}
\end{center}
\end{figure}

In $f(R)$ models, the Einstein-Hilbert action is supplemented by the addition of a
 free function of the
Ricci scalar $R$.  The critical property of these models is the existence of an extra scalar
degree of freedom $f_R = df/dR$ and the inverse-mass or Compton scale associated with it.  
The square of this length in units of the Hubble length is proportional to
\begin{eqnarray}
B&=& {f_{RR} \over 1+f_R} {R'}{H \over H'} \,,
\end{eqnarray}
where $f_{RR}= d^2 f/dR^2$.
Below the Compton scale, the metric ratio $g \rightarrow -1/3$.

\begin{figure}[tb]
\begin{center}
\includegraphics[width=3.4in]{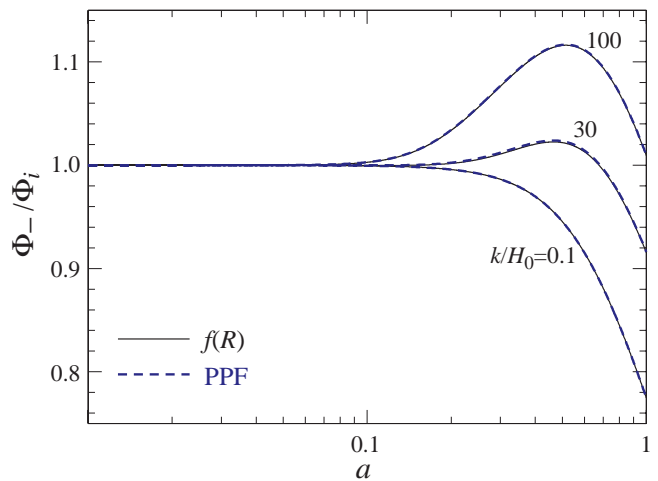}
\caption{Evolution and scale dependence of $\Phi_-$ in $f(R)$ models compared
with the PPF fit.   Here $B_{0}=0.4$, $w_{\rm eff}=-1$ and $\Omega_{m}=0.24$.} 
\label{fig:phim_fr}
\end{center}
\end{figure}

  The evolution of $B$ and the expansion history come
from solving the modified Friedmann equation obtained by varying the action with 
respect to the metric.  We follow the parameterized approach of \cite{SonHuSaw06} where
a choice of the expansion history through 
 $w_{\rm eff}$ and the Compton scale today  $B_0 \equiv B(\ln a=0)$
 implicitly describes the $f(R)$ function and model.   For illustrative
purposes, we take $\Omega_m=0.24$ and $w_{\rm eff}=-1$.

Given $H(\ln a)$ and $B(\ln a)$, the metric ratio at super-horizon scales comes from
solving Eqn.~(\ref{eqn:bertschinger}) 
\begin{eqnarray}
&&\Phi'' + \left( 1 - {H'' \over H'} + {B' \over 1-B} + B {H' \over H} \right) \Phi'\\
&&\qquad  + \left( {H' \over H} - {H'' \over H'} + {B' \over 1-B} \right) \Phi =0 \,,\,\,\, (\kh \rightarrow 0)\,. \nonumber
\label{eqn:Bertclosed}
\end{eqnarray}
We have used the $f(R)$ relation \cite{SonHuSaw06}
\begin{equation}
\Phi + \Psi = -B { H' \over H}{V_m \over \kh} \,,
\end{equation}
which when combined with $\zeta'=0$ and Eqn.~(\ref{eqn:zetaprimedefn}) gives
\begin{equation}
\Psi = {-\Phi - B \Phi' \over 1-B} \,,\,\,\, (\kh \rightarrow 0)\,.
\label{eqn:frpsi}
\end{equation}
The solution of Eqn.~(\ref{eqn:Bertclosed}) together with (\ref{eqn:frpsi}) yields
the metric ratio
\begin{equation} 
g(\ln a, \kh=0)= g_{\rm SH}(\ln a) = {\Phi+\Psi \over \Phi - \Psi}  \,.
\end{equation}
The density evolution function $f_\zeta$ can be adequately described by noting that
$\zeta' \propto \kh^2 B\zeta$ and that $B$ also controls the behavior of $g$.   We take
\begin{equation}
f_\zeta = c_\zeta g
\label{eqn:fzeta}
\end{equation}
with $c_\zeta\approx-1/3$. 

For the transition to the quasi-static regime we take the interpolating function 
\begin{align}
 g(\ln a,k)  &= { g_{\rm SH} +  g_{\rm QS}(c_{g}\kh)^{n_{g}} \over 1+ (c_{g}\kh)^{n_{g}}}\,,
 \label{eqn:ginterpolation}
\end{align}
where $g_{\rm QS}=-1/3$.   We find that the evolution is well described by 
$c_g=0.71 B^{1/2}$ and $n_g=2$.   We show an example of this fit
in Fig.~\ref{fig:gfit_fr}.

Finally, the effective Newton constant is rescaled
by $f_R$ and the quasi-static transition takes place near the horizon scale
\begin{equation}
f_G = f_R \,, \qquad c_\Gamma=1 \,.
\end{equation} 
In Fig.~\ref{fig:phim_fr}, we show how well the PPF parameterization reproduces the
full $f(R)$ metric evolution for scales that span the Compton wavelength transition in a 
$w_{\rm eff}=-1$ and $B_0=0.4$ model.  We have checked that a wide range of $f(R)$
models including those of \cite{HuSaw07}
produce comparable matches with these parameter choices.

\subsection{DGP Model}
\label{sec:dgplinear}

In the DGP braneworld model, the transition between two different behaviors for $g$ occurs
at the horizon scale.  Above the horizon, the propagation of 
perturbations into the bulk requires solving the full 5D perturbation equations \cite{deffayet02b}.
Fortunately, above the horizon scale the evolution is scale free and can be solved using the
iterative scaling method of \cite{SawSonHu06}.  Well below the horizon the evolution reaches
the quasi-static limit where the equations can be effectively closed on the brane \cite{LueScoSta04,KoyMaa06}. 
 We therefore take a similar approach to the $f(R)$ case of interpolating between these two
 well-defined regimes.
 
 \begin{figure}[tb]
\begin{center}
\includegraphics[width=3.4in]{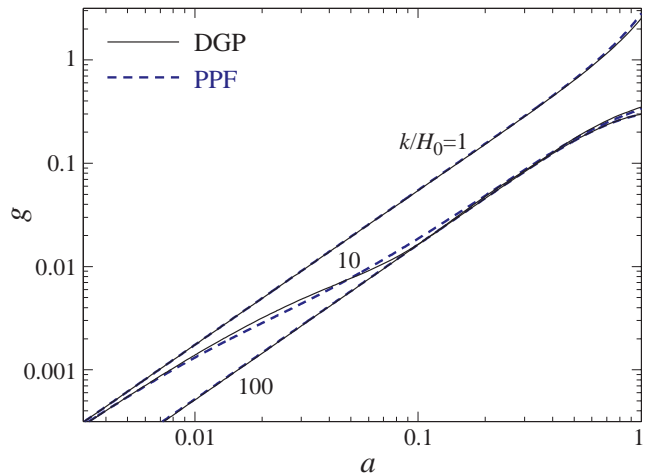}
\caption{Evolution and scale dependence of the metric ratio $g$ in the DGP model compared
with a PPF fit. Here $\Omega_{m}=0.24$ and the PPF parameter $c_g=0.4$.} \label{fig:gfit_dgp}
\end{center}
\end{figure}

 On  super-horizon scales, the iterative scaling solution is well described by
 the fitting function
 \begin{equation}
  g_{\rm SH}(\ln a)={ 9 \over 8 H r_c -1} \left( 1 + {0.51 \over H r_c - 1.08} \right) \,,
\end{equation}
where recall that $r_c$ is the crossover scale.  In the DGP model $\zeta'$ is again related to
$g$ and so we take $f_\zeta$ to be defined by Eqn.~(\ref{eqn:fzeta}) with $c_\zeta \approx 0.4$.
The expansion history is given by 
\begin{align}
{H \over H_0} 
&=
\sqrt{\Omega_\text{rc}}+\sqrt{\Omega_\text{rc}+\Omega_{
m}a^{-3} }  \,,
\label{eqn:dgpexpansion}
\end{align}
where
\begin{align}
\Omega_\text{rc} & ={1 \over 4r_\text{c}^2H_0^2 } = 
\frac{(1-\Omega_{m})^2}{4}.
\end{align}
For illustrative purposes we take $\Omega_m=0.24$.

\begin{figure}[tb]
\begin{center}
\includegraphics[width=3.4in]{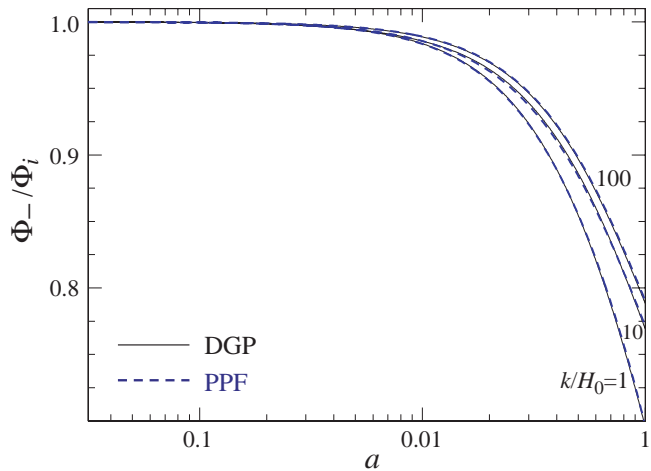}
\caption{Evolution and scale dependence of $\Phi_-$ in the DGP models compared
a PPF fit. Here $\Omega_{m}=0.24$ and the PPF parameter $c_g=0.14$.} 
\label{fig:phim_dgp}
\end{center}
\end{figure}

In the quasi-static regime \cite{KoyMaa06}
\begin{align}
 g_{\rm QS}(\ln a) &= -{1\over 3}\left[ 1-2 H r_{c}\left( 1 + {1 \over 3}{H' \over H}\right) \right]^{-1}\,.
\end{align}
We employ the interpolation function  (\ref{eqn:ginterpolation}) to join the two regimes.
In Fig.~\ref{fig:gfit_dgp}, we show a fit to the results of \cite{SawSonHu06} with
$c_g=0.4$ and $n_g=3$ for several values of $\kh$ that span the transition.
The remaining parameters are
\begin{equation}
f_G = 0 \,, \qquad c_\Gamma=1 \,.
\end{equation} 

Around horizon crossing, the scaling assumption of \cite{SawSonHu06} is briefly violated
leading to possible numerical transients in the solution and an ambiguity in the exact value of $c_g$.
The results for the metric evolution in \cite{SawSonHu06} are best fit with $c_g=0.14$
as shown in Fig.~\ref{fig:phim_dgp}.  The transition parameter $c_g$ should therefore
be taken as a free parameter in the range $c_g \sim 0.1-1$ until a more precise solution is
obtained.

\section{Non-linear Parameterization}
\label{sec:nonlinear}

As discussed in \S \ref{sec:nonlinearregime}, we expect that a successful modification
of gravity will have a non-linear mechanism that suppresses modifications within 
dark matter halos.  In this section, we construct a non-linear PPF framework based on the
halo model of non-linear clustering.  Although a complete parameterized
description of modified gravity
in the non-linear regime is beyond the scope of this work, the halo model framework allows
us to incorporate the main qualitative features expected in these models.  
Searching for these qualitative features can act as a first step for cosmological
tests of gravity in the non-linear regime.

 \begin{figure}[tb]
\begin{center}
\includegraphics[width=3.4in]{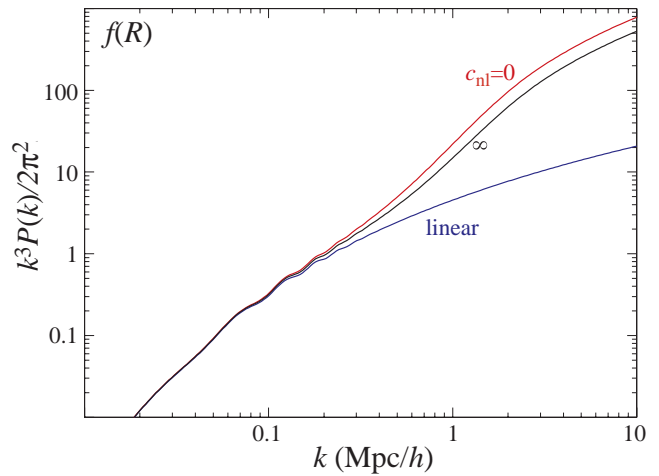}
\caption{PPF non-linear power spectrum ansatz for an $f(R)$ model. 
The non-linear power spectrum is constrained to lie between two 
extremes:  defined by halo-model mass functions with the quasi-static
growth rate [$c_{\rm nl}=0$ or $P_{0}(k)$]
 and the smooth dark energy growth rate with the same expansion history
 [$c_{\rm nl}=\infty$ or $P_{\infty}(k)$]. 
 Here $B_{0}=0.001$, $w_{\rm eff}=-1$ and $\Omega_{m}=0.24$ with other parameters
given in the text.  
}
\label{fig:pk_fr}
\end{center}
\end{figure}

Under the halo model, the non-linear matter power spectrum is
composed of two pieces (see \cite{CooShe02} for details and a review).   One piece involves the correlations between dark matter halos.
As in general relativity, the interactions between halos should be well described by 
linear theory.  The other piece involves the correlations within dark matter haloes. 
  It is this term that we mainly seek to parameterize.

Specifically
given a linear power spectrum of density fluctuations $P_L$, the halo model defines the
non-linear spectrum as the sum of the one and two halo pieces
\begin{align}
P(k) &= I_{1}(k) +  I_{2}^2(k) P_L(k) \,,
\label{eqn:pkhalo}
\end{align}
with
\begin{align}
I_{1}(k) &= \int {d M \over M} \left( { M \over \rho_{0} } \right)^2
	\left[ {d n\over d\ln M} y^2(M,k) \right] \,, \nonumber\\
I_{2}(k) &= \int {d M \over M} \left( { M \over \rho_{0} } \right) 
		{dn \over d\ln M}
		b(M) y(M,k)\,,
	\end{align}
where $\rho_{0} = \rho_m(\ln a=0)$.
Here the integrals are over the mass $M$ of dark matter halos and $dn/d\ln M$ is the mass function which describes the comoving number density of haloes.   $y(M,k)$ is the Fourier transform of
the halo density profile normalized to $y(M,0)=1$ and $b(M)$ is the halo bias.   Note that
$I_2(k=0)=1$ so that the linear power spectrum is recovered on scales that are larger
than the extent of the halos.

 \begin{figure}[tb]
\begin{center}
\includegraphics[width=3.4in]{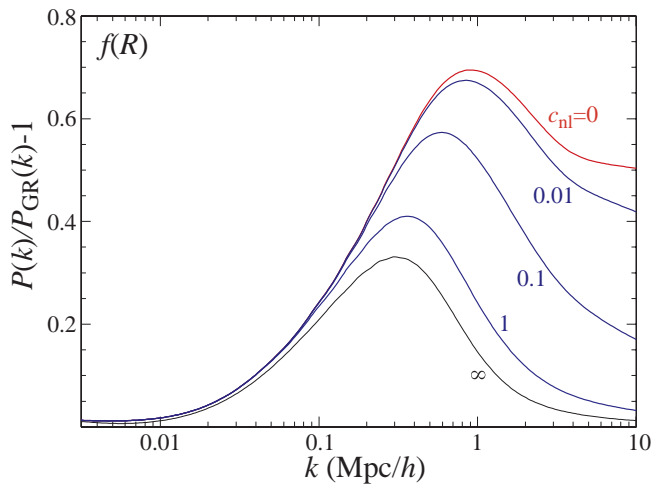}
\caption{Fractional difference in $P(k)$ of the PPF non-linear $f(R)$ ansatz from the smooth dark energy prediction
with the same expansion history.
As $c_{\rm nl} \rightarrow \infty$ deviations become confined to the weakly non-linear to
linear regime.  The model parameters are the same as in Fig.~\ref{fig:pk_fr}.}
\label{fig:nonlin_fr}
\end{center}
\end{figure}

A simple ansatz that restores general relativity in the non-linear regime is that
the mass function and halo profiles remain unchanged from general relativity.  
Specifically, whereas the mass function and halo profiles usually depends in a universal
manner on $\sigma(M)$ 
the 
rms of the linear density field smoothed on a scale that encloses the 
mass $M$ at the background density, we replace this with the rms of
the linear density field of a smooth dark energy model with the same
expansion history $\sigma_{\rm GR}(M)$.   For definiteness,
we adopt the Sheth-Torman mass function and bias \cite{SheTor99}
\begin{align}
{d n \over d\ln M} &= {\rho_{0} \over M} f(\nu) {d\nu \over d\ln M}\,, \nonumber\\
b(M)& = 1 + {a \nu^2 -1 \over \delta_c}
         + { 2 p \over \delta_c [ 1 + (a \nu^2)^p]}\,,
         \label{eqn:massfn}
\end{align}
where $\nu = \delta_c/\sigma(M)$ and 
\begin{eqnarray}
\nu f(\nu) = A\sqrt{{2 \over \pi} a\nu^2 } [1+(a\nu^2)^{-p}] \exp[-a\nu^2/2]\,.
\end{eqnarray}
We choose
$\delta_c=1.68$, $a=0.75$, $p=0.3$, and $A$ such that $\int d\nu f(\nu)=1$.
For the halo profiles we take the
Navarro, Frenk and White (NFW)
profile \cite{NavFreWhi97}
\begin{equation}
\rho  \propto {1 \over  c r/r_{\rm vir}(1+  c r/r_{\rm vir})^2}\,,
\end{equation}
where $r_{\rm vir}$ is the virial radius,
 \cite{Buletal01}
 \begin{equation}
c(M_v) = {9 \over {1+z} } \left( {M \over M_*} \right)^{-0.13}\,,
\label{eqn:concentration}
\end{equation}
 and $M_*$ is defined as $\sigma(M_*)=\delta_c$.
 We call the result of taking $\sigma(M) = \sigma_{\rm GR}(M)$ in
 the halo model equations (\ref{eqn:pkhalo})  $P_\infty(k)$.

  At the opposite extreme,
we can make the ansatz that the usual mapping of the linear to nonlinear power
found under general relativity remains unchanged in modified gravity.  In this
case, changes in the linear growth rate determine the non-linear power spectrum.
 This type of prescription of adopting
the linear to nonlinear scaling of general relativity has been tested
against cosmological simulations with various modified Poisson prescriptions 
\cite{StaJai06}.
It represents the case where gravity is modified down to the smallest cosmological
scales.   Specifically, in our halo model we take $\sigma(M)=\sigma_{\rm PPF}(M)$ as calculated from the linear power
spectrum of the modified gravity model and employ the same Sheth-Torman and NFW prescriptions
as before.    Let us call the power spectrum in this
limit $P_0(k)$.
 
  \begin{figure}[tb]
\begin{center}
\includegraphics[width=3.4in]{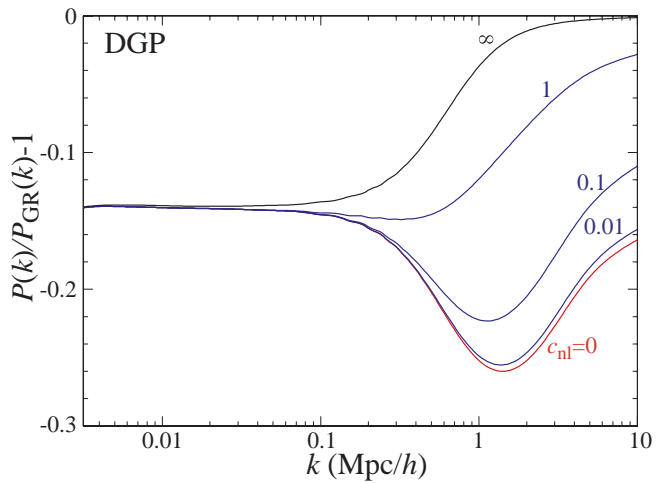}
\caption{Fractional difference in $P(k)$ of the PPF non-linear DGP ansatz from the smooth dark energy prediction.
The DGP model has less power than the equivalent dark energy model in the linear
and weakly non-linear regime.  
Here $\Omega_{m}=0.24$ and other parameters are given in the text.}
\label{fig:nonlin_dgp}
\end{center}
\end{figure}

We can parameterize an interpolation between these two extreme behaviors
\begin{equation}
P(k) = {P_{0}(k) + c_{\rm nl} \Sigma^2(k) P_{\infty}(k)  \over 1 + c_{\rm nl} \Sigma^2(k)}\,,
\end{equation}
that is based on the degree of non-linearity defined by 
\begin{equation}
\Sigma^2(k) \equiv {k^3 P_L(k) \over 2\pi^2} \,.
\end{equation}
The analogous interpolation can also be used for the power spectrum of $\Phi_-$
that enters into gravitational lensing observables \cite{WhiKoc01}.

We show an example of this non-linear ansatz for an $f(R)$ model with 
$\Omega_m=0.24$, $\Omega_m h^2 =0.128$, $\Omega_b h^2=0.0223$, $n_s=0.958$
and initial curvature fluctuation of $\delta_\zeta = A_S^{1/2}= 4.52 \times 10^{-5}$ at
$k=0.05$ Mpc$^{-1}$.
We furthermore fix the expansion history with $w_{\rm eff=-1}$.

Unfortunately no
cosmological simulations exist for $f(R)$ or DGP models against which to test the
accuracy of this non-linear ansatz.   
Moreover
even under general relativity, the halo model model of Eqn.~(\ref{eqn:pkhalo}) does not
exactly reproduce the non-linear spectra of cosmological simulations.  

More robust in our parameterization is the relative change between a PPF power
spectrum and the  general relativistic
prediction with smooth dark energy $P_{\rm GR}(k)$ and the same expansion history.
This factor can then be applied to more exact results from cosmological simulations
to search for deviations from general relativity.
The difference between $P_{\rm GR}(k)$ and $P_\infty(k)$ is that the former uses the
general relativistic linear power spectrum in Eqn.~(\ref{eqn:pkhalo}).    This prescription
can be further refined by calibrating $P_0(k)$ directly from simulations of
the modified Poisson equation \cite{StaJai06}.

We show the fractional change between the PPF power spectra and the general relativistic
power spectra for the $f(R)$ model in Fig.~\ref{fig:nonlin_fr} and a DGP model with the
same parameters but with the DGP expansion history of Eqn.~(\ref{eqn:dgpexpansion})
in Fig.~\ref{fig:nonlin_dgp}.    Note that as $c_{\rm nl} \rightarrow \infty$, deviations
appear mainly in the linear to weakly non-linear regime.  For the $f(R)$ model they
appear as an enhancement of power and for the DGP model as a deficit of power
reflecting the opposite sign of $g$ in the linear regime of the two models.

 \section{Discussion}

We have introduced a parameterized framework for considering scalar modifications
to gravity that accelerate the expansion without dark energy.    This framework
features compatibility in the evolution of structure 
with a background expansion history on large scales, a modification of the
Poisson equation on intermediate scales, and a return to general relativity 
within collapsed dark matter halos.  This return to general relativity is
 required of models to pass stringent local tests of gravity.  
We have also clarified the formal relationship between modified gravity and dark energy in the
Appendix.
A metric based
 modified gravity model can always be cast in terms of a dark energy component with a 
stress energy tensor defined to match its influence on the metric.   However such a component
would possess dynamics which are coupled to the matter.  

Our parameterized post-Friedmann framework features several free functions even in 
the linear regime.  The most important function is the relationship $g = (\Phi+\Psi)/(\Phi-\Psi)$ between
the time-time and space-space pieces of the metric in Newtonian gauge.  
Supplementing these are two functions that link the metric to matter density perturbations:
one on super-horizon scales and one on intermediate scales.   Finally there is a parameter
that controls the interpolation between these two regimes.   

We have shown that with an
appropriate choice of parameters this
framework describes linear perturbations in 
the $f(R)$ modified action and DGP braneworld gravity models.   It may be used
in place of the more complicated 4th order and higher dimensional dynamics exhibited
in these models respectively when studying phenomena such as the integrated Sachs-Wolfe
effect in the CMB, large-scale gravitational lensing
 and galaxy clustering.  We intend to explore
these applications in a future work.

On non-linear scales our framework features an ansatz based on the requirement that
scalar modifications should be suppressed locally in order to pass the stringent tests
of general relativity in the solar system.   Indeed the scalar degrees of freedom in
both the $f(R)$ and the DGP models possess non-linearities that
drive the dynamics back to general relativity in high curvature or high density regimes.  
Our ansatz is based on the halo model of non-linear clustering.  It allows for a
density dependent interpolation for the abundance and structure of dark matter halos
between the expectations of general relativity and
the modified Poisson equation on 
intermediate scales.

Due to the current lack of cosmological simulations in these modified gravity models,
the accuracy of our simple ansatz remains untested.    With cosmological simulations,
our framework can be extended and refined by introducing more parameters that 
describe the potentially mass-dependent modification of dark matter haloes.   
In fact, our simple halo model parameterization is not even sufficient to accurately model non-linear
effects in general relativity. 
Nonetheless phrased as a simple template form for relative deviations in the
power spectrum between modified gravity and general relativity with smooth dark
energy, our current ansatz can be used in conjunction 
with more accurate results from dark energy cosmological simulations.  For
example, it can be used to search for possible deviations of this type  as
a consistency check on dark energy inferences from expansion history tests with
upcoming cosmic shear surveys.

While many such consistency tests have been proposed in the literature, 
it is important to incorporate a density dependence to the 
modifications as we have done here.  
The principle 
that non-linear scales should exhibit a return to general relativity itself suggests
that mildly non-linear scales provide the most fruitful window for cosmological
tests of gravity. Furthermore uncertainties 
in the baryonic influence on the internal structure of dark matter halos in the deeply
non-linear regime even under
general relativity
(e.g.~\cite{Jinetal06,RudZenKra07})
make consistency tests in this regime potentially
ambiguous.  Our parameterized framework should enable studies of such issues
in the future.

\appendix

\section{Dark Energy Correspondence}

Suppose we view the modifications to gravity
in terms of an additional ``dark energy" stress  tensor.  
We are free to {\it define} the dark energy stress tensor to be 
\begin{equation}
T^{\mu\nu}_{e} \equiv {1 \over 8\pi G} G^{\mu\nu} - T^{\mu\nu}_{m} \,.
\label{eqn:effectivede}
\end{equation}
Given this association, all of the familiar structure of cosmological 
perturbation theory in general relativity applies.  In particular,
covariant conservation of the matter stress energy tensor $T^{\mu\nu}_m$
and the Bianchi identities imply conservation of the effective dark energy 
\cite{Bashinsky:2007yc}
\begin{equation}
\nabla_\mu T_e^{\mu\nu} =0 \,.
\end{equation}
The remaining degrees of freedom in the effective dark energy stress tensor can then 
be parameterized in the same manner as a general dark energy component 
\cite{Hu98,Kunz:2006ca}.  Two models that imply the same stress energy
tensor of the combined dark matter and dark energy at all points in
spacetime are formally indistinguishable gravitationally \cite{HuEis99,GiaHu05}.

Note however that this equivalence is only formal and two physically distinct models,
e.g. $f(R)$ modified gravity and scalar field dark energy, will not in general imply the
same effective stress energy tensor  \cite{Albrecht:2006um}.  
The Einstein and conservation
equations do not form a closed system and the distinction between modified
gravity and dark energy lies in the closure relation.   For dark energy that is
not coupled to matter, the closure relationship takes the form of equations of
state that define its internal dynamics.  These micro-physical relations do not
depend explicitly on the matter.  
For example for scalar field dark energy, the sound speed or the
relationship between the pressure and energy density fluctuations is defined in
the  constant field gauge without reference to the matter
\cite{Hu98}, and is associated with the form of the
kinetic term in the Lagrangian \cite{GarMuk99}.

For modified gravity of the type described in this paper, we shall
see that the closure relations must depend explicitly on the matter
(see also \cite{Bashinsky:2007yc}).  
The effective dark energy of a modified gravity model must
be coupled to the matter.  In other words, while the modification of gravity can be modeled
as fifth forces mediated by the effective dark energy, it cannot be viewed
as a missing energy component that obeys separate equations of motion.

It is nonetheless useful to phrase the PPF parameterization in terms of
an effective dark energy component.   It enables the use of the extensive
tools developed for cosmological perturbation theory and facilitates 
the development of PPF formalisms in different  gauges.

\subsection{Covariant Field and Conservation Equations}

Following \cite{Bar80,KodSas84}, we parameterize linear scalar metric fluctuations
of a comoving wavenumber $k$ as 
\begin{align}
{g^{00}} &= -a^{-2}(1-2 {\potential}Y)\,, \nonumber\\
{g^{0i}} &= -a^{-2} {\shift} Y^i\,,  \nonumber\\
{g^{ij}} &= a^{-2} (\gamma^{ij} -
        2 {\curvature} Y \gamma^{ij} - 2 {\shear Y^{ij}})\,,
\label{eqn:metric}
\end{align}
where the $``0"$ component denotes conformal time $\eta =\int dt/a$ and
 $\gamma_{ij}$ is the background spatial metric which we assume to
be flat across scales comparable to the wavelength.  Under this assumption,
the spatial harmonics are simply plane waves
\begin{align}
Y & = e^{i {\bf k}\cdot {\bf x}}\,, \nonumber\\
Y_i &  = (-k) \nabla_i Y \,,\nonumber\\ 
Y_{ij} &= (k^{-2} \nabla_i \nabla_j + \gamma_{ij}/3) Y \,.
\end{align}
Likewise the components of the stress tensors can be parameterized as
\begin{align}
{T^0_{\hphantom{0}0}} &= -\rho - { \delta\rho}\,, \nonumber\\
{T_0^{\hphantom{i}i}} &= -(\rho + p){v}Y^i\,, \nonumber\\
{T^i_{\hphantom{i}j}} &= (p + {\delta p}Y)  \delta^i_{\hphantom{i}j} 
	+ p{\Pi Y^i_{\hphantom{i}j}}\,, 
\label{eqn:stressenergy}
\end{align}
where we will use the subscripts $m$ to denote the matter and $e$ to
denote the effective dark energy.  When no subscript is specified we
mean the components of the total or matter plus effective dark energy 
stress tensor. For simplicity we assume that the
radiation is negligible during the epochs of interest.

By definition, Eqn.~(\ref{eqn:effectivede}) enforces the usual 4 Einstein
field equations \cite{HuEis99}
\begin{align}
& {H_L}+ {1 \over 3} {H_T}+   {B \over \kh}-  {H_T' \over \kh^2} 
   \nonumber\\&\qquad
   = { 4\pi G \over H^2 \kh^{2}}   \left[ {\delta \rho} + 3  (\rho+p){{v}-
{B} \over \kh }\right] \,,
\nonumber\\
&  {A} + {H_L} +  {H_T \over 3} + 
 {B'+2B \over \kh} - 
 \left[ {H_T'' \over \kh^2} + \left( 3 + {H'\over H}\right) {H_T' \over \kh^2} \right]
   \nonumber\\&\qquad
 = -{8\pi G \over H^2 \kh^2} {p\Pi}  \,,
\nonumber\\
& {A} - {H_L'} 
- { H_T' \over 3} 
   =  {4\pi G \over H^2 } (\rho+p){{v}-{B} \over \kh} \,,
\nonumber\\
 & A' + \left( 2 + 2{H' \over H} - {\kh^2 \over 3}  \right) A 
-{\kh \over 3}  (B'+B) \nonumber\\&\qquad - H_L'' - \left( 2 + {H' \over H}\right) H_L' 
   = {4\pi G \over H^2} ({\delta p} + {1 \over 3}{\delta\rho} ) \,,
\label{eqn:einstein}
\end{align}
where recall $'=d/d\ln a$ and $\kh = (k/aH)$.   The conservation laws
for the matter and effective dark energy become
\begin{align}
& {\delta\rho'}
	+  3({\delta \rho}+ {\delta p})
=
        -(\rho+p)(\kh {v} + 3 H_L')\,, \\
&      {[a^4(\rho + p)({{v}-{B}})]' \over a^4\kh}
= 
 { \delta p }- {2 \over 3}p {\Pi} + (\rho+ p) {A} \,. \nonumber
  \label{eqn:conservation}
\end{align}
There are 4 metric variables and 4 matter variables per component
that obey 4 Einstein equations and 2 conservation equations per component.  
However 2 out of 4 of the Einstein equations are redundant since the
Bianchi identities are automatically satisfied given a metric.  Furthermore,
2 degrees of freedom simply represent gauge or coordinate freedom.  
This leaves 2 degrees of freedom per component to be specified.
Usually, this involves defining equations of state that specify the spatial
stresses in terms of the energy density and velocities.  As we shall
see, it is this prescription that must be altered to describe modified
gravity.

\subsection{Gauge}

The scalar gauge degrees of freedom are fixed by gauge conditions.   Under
a gauge transformation defined by the change in conformal time slicing $T$
and spatial coordinates $L$
\begin{align}
 \eta &=\tilde \eta +{T}\,, \\
x^i   &= \tilde x^i +{L}Y^i\,, \nonumber
\end{align}
the metric  variables transform as
\begin{align}
 A &=\tilde A - aH (T'+T)  \,, \nonumber\\
 B &=  \tilde B +  aH( L' + k_H {T}) \,, \nonumber\\
 H_L &=  \tilde H_L - aH(T + {1\over 3}k_H) \,, \nonumber\\
 H_T &=  \tilde H_T + aH k_H  {L}\,, 
\label{eqn:metrictrans}
\end{align}
and the matter variables transform as
\begin{align} 
{\delta \rho} &= \widetilde{\delta\rho} - \rho' aH {T}\,, \nonumber\\ 
{\delta  p} &= \widetilde{\delta p} -  p' aH  {T}\,, \nonumber\\
 v &=  \tilde v +   aH { L'} \,, \nonumber\\
 \Pi &= \tilde \Pi \,.
\label{eqn:fluidtrans}
\end{align}
A gauge is fully specified if the functions $T$ and $L$ are uniquely defined.

In this paper we work in the matter comoving and Newtonian gauges.  
The matter comoving gauge is specified by the conditions
\begin{align}
B &= v_m \,, \nonumber\\
H_T &= 0 \,.
\end{align} 
They fully specify the gauge transformation from an alternate gauge
choice
\begin{align}
T &= (\tilde v_m - B)/k  \,, \nonumber\\
L &=  -\tilde H_T /k \,.
\end{align}
To avoid confusion between fluctuations defined in different gauges, we will define
\begin{eqnarray}
\zeta &\equiv& H_L \,,\nonumber\\
\xi &\equiv& A  \,,\nonumber\\
 \rho\Delta  &\equiv& \delta \rho \,, \nonumber\\
\Delta p &\equiv& \delta p \,, \nonumber\\
V &\equiv& v \,.
\end{eqnarray}
$\Delta p$ should not be confused with $p \Delta = p (\delta \rho/\rho)$.

The appropriate Einstein and conservation equations for this gauge can be obtained
by utilizing these definitions in Eqn.~(\ref{eqn:einstein}) and (\ref{eqn:conservation}).
For example, the third Einstein equation reads
\begin{equation}
\zeta' = -{4\pi G \over H^2}(\rho_e + p_e) {V_e - V_m \over k_H}\,,
\label{eqn:zetaprimeeff}
\end{equation}
and the energy-momentum conservation equations  for the matter become
\begin{align}
\Delta_m ' &= -\kh V_m - 3\zeta' \,,\nonumber\\
\xi &= 0 \,.
\label{eqn:energycom}
\end{align}
The dark energy momentum conservation equation
\begin{align} &
    {[a^4(\rho_e + p_e)({{V_e}-{V_m}})]' \over  a^4 \kh}    
= 
 { \Delta p_e }- {2 \over 3}p_e {\Pi_e} \,,
\end{align}
 in conjunction with 
 Eqn.~(\ref{eqn:zetaprimeeff}) and the first Einstein equation 
 implies that unless $\Delta p_e$ or $p_e\Pi_e > {\cal O}(\Delta \rho/\kh^2)$, $\zeta' /\zeta 
 \rightarrow 0$ as $\kh \rightarrow 0$.  

Similarly, the Newtonian
gauge is defined by the condition $B=H_T=0$ and the transformation 
\begin{align}
T &= -{\tilde B \over k} + { \tilde H_T'  \over k \kh } \,, \nonumber\\
L &=  -{\tilde H_T \over k} \,.
\end{align}
To avoid confusion we define
\begin{align}
\Phi &\equiv H_L \,,\nonumber\\
\Psi &\equiv A \,.
\end{align}
We refrain from utilizing  matter variables in Newtonian gauge but note that
velocities in the two gauges are the same.  
The relationship between the two metric fluctuations are 
\begin{align}
\zeta &= \Phi - {V_m \over \kh} \,, \nonumber\\
\xi & = \Psi - { V_m ' + V_m \over \kh} \,.
\label{eqn:comnewt}
\end{align}
The matter momentum conservation law in Newtonian gauge
becomes
\begin{align}
V_m' + V_m = \kh \Psi \,.
\label{eqn:momentumnewt}
\end{align}
This equation can alternately be derived from the gauge transformation equation
(\ref{eqn:comnewt}) given that $\xi =0$.

Finally the Einstein equations  (\ref{eqn:einstein}) in Newtonian gauge imply 
\begin{align}
\label{eqn:phiplusgr}
{\Phi+\Psi \over 2}&= - {4\pi G \over H^2 \kh^2} p \Pi = - {4\pi G \over H^2 \kh^2} p_e \Pi_e \,,\\
{\Phi-\Psi \over 2} &= {4\pi G \over H^2 \kh^2} \left( \rho\Delta + 3 (\rho+p){V-V_{m}\over \kh} + p\Pi \right)
\nonumber\\
&= {4\pi G \over H^2 \kh^2} \Big[ \rho_{m}\Delta_{m} + 
\rho_{e}\Delta_{e}+3 (\rho_{e}+p_{e}){V_{e}-V_{m}\over \kh} \nonumber\\
&\quad+ p_{e}\Pi_{e} \Big] \,,
\label{eqn:phiminusgr}
\end{align} 
where 
we have assumed that the anisotropic stress of the matter is negligible.
A finite metric ratio parameter $g = (\Phi+\Psi)/(\Phi-\Psi)$
is thus associated
with a non-vanishing effective anisotropic stress.

\subsection{PPF correspondence}

The system of equations defined by the field equations and the conservation equations
are incomplete.   To close the system of equations two more conditions must be
required of the effective dark energy.  It is this closure condition that the PPF parameterization
must determine.

Given that the matter has no anisotropic stress, Eqn.~(\ref{eqn:phiplusgr})
defines
the anisotropic stress of the effective dark energy in terms of the metric
 \begin{equation}
 p_e\Pi_e =-{ H^2 \kh^2 \over 4\pi G } g \Phi_- \,,
 \end{equation}
 where recall $\Phi_-=(\Phi-\Psi)/2$.  This
 is the first of two closure relations.

The second closure relation comes from equating Eqn.~(\ref{eqn:phiminusgr}) 
and the modified
Poisson equation (\ref{eqn:gammapoisson})
 \begin{equation}
\rho_e  \Delta_e + 3(\rho_e+p_e){V_e-V_m \over k_H} + p_e\Pi_e =- { k^2 \over 4\pi Ga^2} \Gamma\,.
 \end{equation}
 The PPF equation of motion (\ref{eqn:gammaeom}) for $\Gamma$ is therefore the
 ``equation of state" for the effective dark energy.
 
The conservation laws for the effective dark energy and/or remaining Einstein equations
 then define the other two components $V_e$ and $\Delta p_e$.  For example,
\begin{align}
V_e & = V_m - k_H  {H^2 \over 4\pi G a^2 (\rho_e + p_e) }\zeta' \,, \\
\Delta p_e & = p_e \Delta_e - {1\over 3} \rho_e\Delta_e'  - (\rho_e+p_e) (k_H V_e/3 + \zeta')\,.\nonumber
\end{align}
The modification represented
by this prescription obeys all 4 Einstein equations and both sets of conservation laws.

Unlike the case of a micro-physical candidate for dark energy such as a scalar field, the
closure relations not only cannot be defined as direct relationships between the spatial
stresses and the energy density and velocity, they here involve the matter and the metric
fluctuations directly.  Hence the effective dark energy is implicitly coupled to the matter
and cannot be described as an independent entity.

On the other hand, the virtue of making this correspondence explicit is that with these relations
all of the usual representations of perturbation theory can be reached by standard gauge
transformations from our matter comoving and Newtonian representations.

\vspace{1.cm}

\noindent {\it Acknowledgments}: We thank Dragan Huterer, Bhuvnesh Jain,
Yong-Seon Song, and
 Amol Upadhye for useful conversations.
This work was supported by the
U.S.~Dept. of Energy contract DE-FG02-90ER-40560, 
the David and Lucile Packard Foundation
and the KICP under NSF PHY-0114422. 
\hfill\vfill
\bibliography{ppf}

\end{document}